\documentclass[opinion]{sigirforum}

\def\pubissue{Submitted to ACM SIGIR Forum March 2025.}

\usepackage{wrapfig}

\begin{document}
\title{Widening the Role of Group Recommender Systems with CAJO}

\authors{
\author[francesco.ricci@prof.senior.unibz.it]{Francesco Ricci}{Free University of Bozen-Bolzano}{Bozen-Bolzano, Italy}
\and
\author[adelic@etf.unsa.ba]{Amra Deli\'{c}}{University of Sarajevo}{Sarajevo, Bosna \& Herzegovina}
}

\maketitle 
\begin{abstract}

Group Recommender Systems (GRSs) have been studied and developed for more than twenty years. However, their application and usage has not grown. They can even be labeled as failures, if compared to the very successful and common recommender systems (RSs) used on all the major ecommerce and social platforms. As a result, the RSs that we all use now, are only targeted for individual users, aiming at choosing an item exclusively for themselves; no choice support is provided to groups trying to select a service, a product, an experience, a person, serving equally well all the group members. In this opinion article we discuss why the success of group recommender systems is lagging and we propose a research program unfolding on the analysis and development of new forms of collaboration between humans and intelligent systems. We define a set of roles, named CAJO, that GRSs should play in order to become more useful tools for group decision making.

\end{abstract}

\section{Introduction}



Group Recommender Systems (GRSs) are mirroring  classical Recommender Systems~\citep{2022rsh}, by providing relevant recommendations for products and services (items), when the target is not an individual person, but a group of people that will consume together the chosen item~\citep{MasthoffD22,Felfernig2024-ce}. Common scenarios for GRSs are: helping a person to find music to be played in a gym while many people are training, or advising a family trying to select a destination for a week end holiday. In practice many usage scenarios for GRSs can be created by changing the nature of the group context (size, persistency, reciprocal knowledge, structure, diversity) and the type of item (experience, service, representative person, product)~\citep{JamesonWF22}. 

GRSs researchers have been designing and testing their systems for more than twenty years. They have targeted what was identified as the core technical problem to solve for generating effective group recommendations: individual group members' preference aggregation. Inspired by social choice theory, researchers claimed that proper recommendations for a group must aggregate and balance the preferences of all the group members, and must resolve potential preference conflicts among the group members' goals. While this sounds reasonable and compelling, the reality is that no choice will make all the group members equally happy, since people may be happy in completely different ways, and these multi-criteria decision problems do not have a unique optimal solution~\citep{arrow74}.

Ultimately, we have not seen real (commercial) applications of GRSs growing as quickly and as strongly as more common recommender systems, i.e., those supporting the choices of individual users~\citep{2022rsh}, which are now standard components of most of the large social, media and ecommerce platforms. There is surely a range of reasons why GRSs research has not yet produced effective solutions. In this article we compare the application problem with the conducted research to better understand why the success of group recommender systems is lagging. Overall, we believe that the core application problem, i.e., supporting group decision making~\citep{JamesonWF22}, has not been properly targeted in the mainstream system design, and the research has focused on trying to automatically solve an unsolvable problem instead of helping groups to rationally face and live with such a problem: making everybody happy.

In this opinion paper we propose a research program addressing a novel and wider scope definition of GRSs and their functionality. The proposed program is unfolding on the analysis and the development of new forms of collaboration between humans and artificial intelligent agents. We will define a set of roles that GRSs should implement in order to become a more useful tool for group decision making. These roles can be also seen as artificial intelligence agents able to collaborate with humans. We will refer to some recent work that is moving along these directions and we will discuss future research activities that should be addressed by the research community.
\section{Group Decision Making}
\label{sec:gdm}


Group decision-making is an essential part of everyday life, influencing a wide range of decisions. These can be simple and repetitive, such as, deciding what to have for family dinner, what movie to watch, or what music to play in a household and keep everyone satisfied. However, they can also be more complex and unique, like the decision a committee is called to make by selecting a job candidate, or a company's board members making strategic decisions about the organization's future. Regardless of the context, group decisions require some level of member involvement, including gathering and sharing information, discussing alternatives, collaborating, and compromising, as it was shown by~\citep{forsyth14}.

Group decisions often differ significantly from those made by individuals, even when addressing the same issues. For example, a couple selecting household furniture together may reach a completely different decision if each person were to choose independently. In some cases, group decisions benefit from the diverse knowledge, information, and skills brought by the different members, while in other cases, social interactions can sometimes become distracting, leading to poorer decision outcomes~\citep{StasserD2001}. Generally, groups are better at problem-solving than individuals \citep{Shaw1932}, and they tend to make better decisions overall. However, there are exceptions to this, for instance, \textit{groupthink}, i.e., when a group reaches a consensus without critical reasoning or properly evaluating the consequences and the alternatives~\citep{forsyth14,JamesonWF22}.

\begin{figure}
    \centering
    \includegraphics[width=0.4\linewidth]{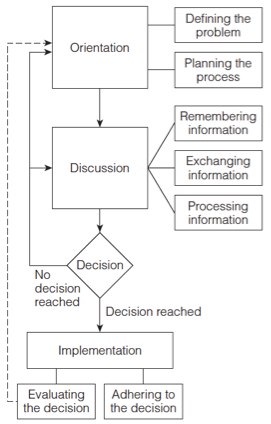}
    \caption{Group dynamics, from \citep{forsyth14}}
    \label{fig:group_dynamics_forsyth}
\end{figure}

According to the functional theory of group decision-making, groups outperform individuals primarily due to their use of structured decision-making procedures. These procedures involve how information is collected, analyzed, and evaluated. While each group decision-making process is unique in terms of how and when these procedures are applied—and no two groups follow exactly the same approach \citep{Delic2018NW}, the theory identifies a common set of decision-making phases, as illustrated in Figure~\ref{fig:group_dynamics_forsyth}.

The first phase, orientation, involves defining the problem—what needs to be decided or the specific decision-making task. The group also determines how to approach the decision, agreeing whether to follow a structured process or not, and establishes a goal that will serve as a basis for evaluating the final outcome.

In the second phase, discussion, group members gather relevant information individually or collectively, recall important details, and exchange this information within the group for collective processing. This discussion leads to the next phase, i.e., decision-making, where the group evaluates different alternatives. A final decision may be reached through consensus, voting, or another decision-making method deemed appropriate. If the group fails to agree, they may revisit the orientation or discussion phase, repeating the process as needed until an acceptable decision is made.

Once the decision is reached, the final phase is implementation. Here, the group acts according to the decision made and evaluates its outcome based on the initial goals set in the orientation phase. \citep{forsyth14} shows that groups who follow this approach are more likely to make good decisions than those who do not. A complementary discussion of alternative patterns used by people to make choices is presented in~\citep{JamesonWF22}. 



\section{Group Decision Making is Hard}
\label{sec:hard}



So, why group decision making is hard and why groups may benefit from various types of decision support functionalities, in the form of digital solutions? We will focus on four dimension that determine the complexity of the application scenario: coordination, information processing, fairness and contextualization. 

{\bf Information exchange and processing} is fundamental in group decision making. A group primary needs information about alternative items, among which the choice will be made. Moreover, it needs the right amount of information, on the right amount of items: too few and too much will cause dissatisfaction and will make harder to make a choice~\citep{schwartz04}. More specifically, each single group member will have a particular amount of information that will suit his/her decision style~\citep{fesenmaier03}, and it will be difficult to create and manage multiple channel of communications, so that each group member will be optimally served~\citep{DelicE0M24}. In addition to item related information, group members need information about the other members (preferences, personality, goals) and about the decision process (already executed and to be complete communication and decision actions). Moreover, people adapt to the other group members, by modifying their preferences~\citep{masthoff06} and conflict resolutions styles~\citep{NguyenRDB19}. Finally, having a clear understanding of the decision problem is fundamental to faithfully implement a rational decision making schema.

{\bf Coordination} of the group decision making process is a strict requirement for a group to meet the chosen decision goals~\citep{TranFL24}. Cooperative and uncooperative behaviors must be sustained and moderated, respectively, in order to achieve high decision quality~\citep{NguyenRDB19}. Who in the group is taking care of this coordination is also critical: social relationships and trust impact onto the decision or the emergence of leadership~\citep{NajafianMKT24}. Moreover, group members may show diverse levels of assertiveness while arguing in favor of their preferred items and against those that do not meet their needs and wants~\citep{THOMAS08}. This assertiveness, e.g., how frequently a group member reaffirm her/his preferences, must be moderated to avoid decision failure. Finally, biases in group decision making are often caused by weak time management, such as loosing excessive amount of time on minor aspects or excessively focusing on individual and divisive preferences~\citep{MasthoffD22}.

Achieving {\bf fairness} in group decision making is hard, if not impossible, as no single computable criterion of fairness exists~\citep{Ekstrand0B022}, and we have already noted that an optimal social choice preference aggregation strategy is not existent~\citep{arrow74}. Notwithstanding its complexity, fairness is a major target for a group, while discussing and making choices~\citep{EmamgholizadehDR24}. Hence, coming up with a viable and accepted definition of fairness is required in any group decision making process. Here, again, individuals will show a range of diverse understanding of what fairness means, in a genuine and manipulative approach, and the group should decide which criteria are to be prioritized~\citep{forsyth14,BarileDIRNFHT24,TranFL24}. This is also related to the coordination dimension: the group is challenged by the decision about how to decide.

Finally, all the above mentioned dimension of complexity are influenced by the {\bf context} of the decision~\citep{AdomaviciusBTU22,JamesonWF22}. The type of the item to be decided, the amount of time available before the decision is to be made, the size of the group, the personality of the group members~\citep{AlvesMSCNM23}, the social relationships between group members, all these contextual factors have a major impact on the final outcome~\citep{MasthoffD22}. A group recommender system must take into account these contextual factors to produce useful recommendations, in the target context. Moreover, the GRS must be highly dynamic to be able to track contextual changes that may require new information processing and coordination actions. 
\section{GRS State of the Art}
\label{sec:grs}

Standard RSs, designed for individual users, suggest items based on an analysis of the target user's collected preferences, and the preferences of a population of other users. Conversely, Group Recommender Systems (GRSs) simultaneously account for the preferences of multiple users, while relying on a data set of collected individual preferences and other groups' choices. A straightforward design choice of GRSs is to accurately represent the preferences of all the group members, and then to identify items that best satisfy the group as a whole~\citep{MasthoffD22}. However, to accomplish this goal, a key challenge, which remains an active research topic, is how to effectively combine (aggregate) individual preferences into a unified group model and/or a set of group recommendations. The methods used for this purpose are known as preference aggregation strategies. However, no single preference aggregation strategy can optimally produce the preferences of all group members, and generate proper recommendations, given the observed variety of contexts and group types~\citep{Delic2018NW}. 


Many of the earliest aggregation strategies are rooted in the Social Choice Theory~\citep{MASTHOFF15}, which studies how individual preferences should be combined into collective decisions (normative approaches). However, according to Arrow's Theorem~\citep{arrow1963}, no single optimal aggregation strategy exists that simultaneously satisfy some very basic fairness criteria. This implies that different strategies can be used, and are effectively used, depending on the group's  context and goals. Common strategies include maximization of the average of individual preference scores for the alternative items, multiplicative scoring (maximizing the product of individual scores), least misery (minimizing dissatisfaction, by taking the item with the largest of the lowest scores given by the group members to the items), maximum pleasure (maximizing satisfaction, by taking the item with the largest of the largest scores), and voting-based methods such as the Copeland rule and Borda count. Each strategy has its advantages and limitations, and by optimizing different criteria, it balances individual preferences in a particular way. For example, the Average strategy ensures fairness by equally considering all preferences, while Least Misery prioritizes items for which individuals do not have low preferences (e.g., respecting a vegetarian that avoids meat, even if this is not optimal for the others).

More advanced aggregation strategies have since been developed, incorporating various techniques to refine group preference modeling. These include minimizing statistical dispersion within the group preference model~\citep{salamo2012}, using negotiation mechanisms to reach a Nash Equilibrium~\citep{carvalho2013}, and aggregating preferences at the attribute level based on Multi-Attribute Utility Theory~\citep{STETTINGER15B}. Other approaches employ distance metrics like Spearman’s footrule~\citep{baltrunas2010group} or leverage random walks over user-item graphs~\citep{kim2013folkommender} to refine recommendations.

Additionally, some reference aggregation strategies take into account social influence and group dynamics. These consider factors such as emotional contagion and conformity~\citep{masthoff06}, activity-based roles \cite{ali2015group}, family-based roles~\citep{BERKOVSKY10}, influence shaped by personality and social trust~\citep{quijano2013}, and patterns of influence from previous group choices~\citep{quintarelli2016}.

More recently, some studies have introduced neural network (NN) architectures to aggregate individual preferences into a group model while also accounting for varying levels of user influence depending on the group and item context~\citep{cao2018attentive,zhang2021double,huang2021novel,yin2019social,sankar2020groupim,guo2021hierarchical}. However, aside from the CAMRa2011 dataset~\citep{said2011challenge}, which provides user-movie preferences along with contextual group data, these models are primarily evaluated on synthetic user check-in data. 
The validation of the quality of the generated group recommendations, in these NN approaches, but also more in general, is very often flawed by the absence of a real ground truth: what the groups decided to choose as good recommendations. In fact, existing data sets almost never include an explicit evaluation of the group for recommendations or the choices (such as the joint group rating for the item). Besides, there are very few data sets with real-group implicit feedback: the chosen items. Furthermore, especially NN approaches are often designed with artificial scenarios in mind, motivated by the available synthetic data, rather than by real-world user-group interactions. 


On the other hand, a smaller body of research has focused on developing complete systems that address various aspects of group recommendations, such as: user-item preference elicitation, user and group modeling, generation, representation and explanations of recommendations, and negotiation support. These systems can be categorized based on their recommendation approach: some of them provide one-time (single-shot) recommendations, while others support an iterative process where preferences are continuously refined over multiple rounds until the group reaches a satisfactory decision. We refer to the latter as conversational GRSs.

Some notable examples of single-shot GRSs include: MUSICFX~\citep{MusicFX98}, which generates a playlist for a gym based on the preferences of the people currently present; PolyLens~\citep{PolyLens01}, which provides movie recommendations for small groups; \citep{yu2006tv} consider TV recommender, which combines user profiles to adapt television programs accordingly; HappyMovie~\citep{quijano2011happymovie}, which recommends movies based on group members' social relationships and personality types; and TourRec~\citep{HERZOG18}, which suggests Point of Interest (POI) itineraries for both individuals and groups. In contrast, conversational GRSs enable more dynamic interactions, such as adapting preferences according to subgroup constraints~\citep{Ardissono03GP}, helping groups decide attributes for a joint vacation~\citep{Jameson04}, facilitating cooperative negotiations~\citep{VenturiniR06}, using critiquing-based techniques~\citep{McCarthy06SCMSN,GuzziRB11}, discussing item features~\citep{MARQUEZ15}, making proposals and counterproposals~\citep{contreras2021integrating}, and even fully chat-based GRSs~\citep{NguyenRDB19}.

In conclusion, it is important to note that decision-making support for groups, as outlined by the functional theory of group decision-making, is still lacking in GRSs and has not been adequately addressed in the research. Additionally, while RSs are widely used in major e-commerce and social-network platforms, GRSs have yet to be implemented in any large-scale application. We believe the reasons for this are multifaceted. First, the systems developed so far were primarily designed to address specific research questions and were never made available for further development by other researchers. Second, people are already making joint decisions by leveraging various communication platforms (e.g., instant messaging), and it seems unlikely that they would move these tasks to alternative dedicated platforms. Both of these issues also explain why there is a lack of real-world datasets capturing group decision-making processes. This is a significant barrier, limiting the development of models and algorithms that could genuinely support groups in decision-making tasks.

\section{CAJO}
\label{sec:cajo}


Building systems that can help a group to reach consensus and make decisions is clearly complex and calls for the exploitation of a range of  approaches and techniques. Hence, without dealing with specific solutions, in this opinion paper, we want here to give a general structure to the solution space, by listing a limited number of functional components, which may act as autonomous agents, playing a role in the decision problem solving. These roles could cooperatively address the group recommendation problem in collaboration with humans. This structure can help to decompose the complexity of the full task, and can contribute to direct the research towards the design of useful and reusable technologies. We call this structure CAJO for: Coach, Arbiter, Judge and Oracle.

\begin{itemize}
    \item{\bf Coach.} The first agent of the proposed framework coaches the group members to become and act as better team players. As we discussed in Section~\ref{sec:gdm}, group dynamics is articulated and context dependent; better decision making outcomes require the group members to agree on the decision making process to follow, and to actually perform the foresight steps. The Coach agent takes therefore the task of informing and training a target group on how to: define the decision problem, agree on the decision process, remember and exchange information that may help to reach consensus and make the final choice. Overall, the Coach helps the group to set and achieve members' goals in a rational way (to reach the selected goals in an effective way). The Coach may also help the group to cope with individual differences, especially related to personality and conflict resolution. In fact, we have learned the importance of individual conflict resolution strategies, adopted by group members (cooperative vs. competitive), in determining the quality of the final choice~\citep{NguyenRDB19}. Ultimately we want to stress that the Coach may need to change the organic group behavior, to improve the quality of the choices.

    \item{\bf Arbiter.} For a group to follow a selected discussion and decision process often requires that one of the group members to act as coordinator. We attribute the coordination set of functionalities to the Arbiter agent. The Arbiter: starts and ends the process; requires that the proper steps of the process are performed, e.g., the group decides how to decide, and are tackled at the right time; ensures that proper time is given to each step; encourages group members to participate to the discussion and moderates those that are taking an excessive space; recalls important information, such as the items that are considered viable options, or the needs and constraints that the group has agreed upon. Overall the Arbiter, of any game, intervenes only when it is needed, to keep the process fair and target oriented.

    \item{\bf Judge.} The third proposed agent is responsible for the final action: choosing one option. The Judge is involved when it is time to stop the discussion and select one of the options, among the alternative suggestions that have been brought into the discussion, either by the GRS (the Oracle, see later) or by a group member. The task of the Judge is to tackle the ``impossible'' problem of aggregating the information generated during the previous steps and suggesting proper decisions, having considered the group members' actions during the process. The information used by the judge comprises a number of diverse types of chunks: the preferences expressed by the group members; their constraints; the evaluations made by the group members for the proposals made by other members (positive vs negative); the actions performed by the group members during the process (support, silence, opposition, cooperativeness, competition). The acquired information is the input for the Judge to propose a final choice. The Judge is going to aggregate these chunks of information and formulate a proposal by using criteria and the consensus model~\citep{TranFL24} that the group has determined as the important one (at the beginning of the dynamics), by leveraging a range of technologies rooted in the current research on fairness~\citep{Ekstrand0B022} and preference aggregation~\citep{MasthoffD22}. We finally note that the Judge is supposed to justify the decision with proper arguments that must include the information and the procedure (e.g. preference aggregation) adopted to determine the choice.

    \item{\bf Oracle.} The last agent that we describe would be cited as the first by many researchers, as it includes the main functionality of current GRSs: suggesting to the group what they should choose. The Oracle is making prediction and responds to any question the group members may have about the process and other group members. The Oracle is typically queried by individual group members to suggest items that the group member may like and can also be accepted by the others; this is a form of one-to-one interaction with the recommender system that was proposed in the past~\citep{GuzziRB11} and recognized as useful by interviewed users~\citep{DelicE0M24}. But the Oracle may also be queried by the group to obtain recommendations that are immediately shown to all, which is the standard approach of many GRSs. Moreover, the Oracle may extend the range of predictions to other aspects. For instance, in a recent study it has been claimed that bringing to the surface the more likely choice that the group is going to make is an important information that is computable and useful to make even better choices~\citep{EmamgholizadehDR24}. In fact, the group can react to this information and reflect on the interactions performed so far, better understanding the state of the discussion and asses the need to deviate from the current direction. The Oracle may also give factual information such as: who is going to be unsatisfied by one option; who is more satisfied by another option; who has been more active or less active in the group dynamics; who has provided more information and who has possibly tried to manipulate the group towards a specific choice. 
\end{itemize}

\section{Works in Progress}

Computer-mediated and computer-supported group interactions have been extensively studied for decades~\citep{mcgrath1998groups,nunamaker1996lessons}, whether through mailing systems~\citep{flores1988computer} or in-person meetings equipped with workstations, large-screen electronic blackboards, and other audiovisual support, to facilitate session planning, group interaction, organizational memory, individual work, and data collection~\citep{nunamaker1991electronic}. However, research on supporting task-oriented, conversational group interactions remains extremely limited, especially considering the variety and complexity of the information support, as it was structured in the previous section. We now point out some recent studies that have addressed one or more functional components of the proposed CAJO structure.

The most influential work for our proposal of CAJO structure is the ARCADES model~\citep{JamesonBGCGVR14,JamesonWF22}. Six strategies are proposed to support user's choices in RSs and GRSs: access information and experience; represent the choice situation; combine and compute; advise about processing; design the domain; evaluate on behalf of the chooser; support communication. All of them can be seen as specialized Oracle functionalities, since they can be used by the group to obtain information about the state of the decision process and the possible choices.

In fact, most of the work in the research field of Group Recommender Systems, as discussed in Section~\ref{sec:grs}, has focused on the Oracle role, and in particular, generating item recommendations for the individuals within a group, as well as for the group as a whole. In a novel chat-based environment, the Oracle role has been paired with the Judge one. In fact, STSGroup~\citep{STSGroup2018NR} offers both recommendations (Oracle) and choice suggestions (Judge). The implementation of these two functionalities is highly dynamic, i.e., they are considering the users' (explicit) evaluations for items, made within the group discussion, i.e., item related preferences (both at the individual and group levels). But, they are also incorporating contextual information, such as group members’ activity during discussions and their moods, which further refine the recommendations and could be used to implement other Oracle functionalities, such as the group choice prediction.

A more recent system~\citep{CHARM24DER} have tried to generalize the approach used in STSGroup, by offering a domain independent platform for group discussion and recommendations. The platform is called CHARM, and provides top-n group recommendations based on explicit feedback from group members regarding items suggested during the discussion. Notably, it does not rely on any database of items and ratings/reactions for the items (which is the standard approach of most of the RSs). The system exploits only the information brought by the users in the discussion and can therefore be applied to any GRS application domain.

Beyond the recommendation task, a few studies have extended the Oracle role, and implemented the Coach and Arbiter roles in the group discussion process. In MUCA~\citep{MUCA2024}, the system supports the group by: Summarizing discussions and providing insights into user opinions and preferences; Encouraging equal participation from all group members; Introducing relevant subtopics to enrich the conversation; Assisting in conflict resolution to help users reach a consensus efficiently. Similarly, \citep{kim2020bot} proposed an AI agent that arbitrates, coaches and responds to user's queries (Oracle) in a group discussions by: Enforcing time constraints and providing alerts when time is running out; Promoting equal contributions from all members; and Organizing individual opinions and summarizing the overall discussion. While both of these approaches offer sophisticated group decision-making assistance, they mostly focus on arbitration. Specifically, they only marginally coach group members on how to navigate the decision-making process more effectively, but they do use the collected information to generate better recommendations, to respond to important questions the group members may have about the process, and they do not suggest any group decisions (Judge).

A complementary perspective is taken by~ \citep{wagner2022comparing} by introducing an AI agent responsible for finding and proposing an optimal time slot and location for a meeting. This is achieved by analyzing individual calendars and availabilities of group members. The agent not only considers individual proposals but also cross-references them with calendar data, effectively serving the group in the Judge role. 

Finally, although the significance of a structured group decision-making process has been widely recognized in the literature~\citep{mark1999meeting, nunamaker1991electronic, forsyth14}, newly developed intelligent agents that support group work in chat environments have yet to offer a reasonably developed support as we detail in the Coach role. It is worthwhile to note that  early research from the 2000s~\citep{farnham2000structured} introduced a scripting approach, instructing group members on what to discuss at specific stages of the decision-making process. While this method did not explicitly train or inform groups on becoming better decision-makers, it provided a structured workflow that they followed step by step. Despite its simplicity, this approach demonstrated superiority in enabling groups to reach consensus more effectively.

\section{Conclusions and Future Work}


In this opinion paper we have summarized the application problem addressed by Group Recommender systems. We have discussed a wider scope of group decision making and recommendation that the research should address, and have briefly surveyed major contributions to the field. Broadening the view of what a GRS should do has been discussed by other scholars~\citep{JamesonWF22}, and we believe that the marginal impact of the GRss research, as a subfield of RSs, is caused by not having addressed the challenges posed by the implementation of systems that fully support group discussion and decision making. As a consequence, the field has not grown enough to produce compelling results, both in understanding how artificial systems can fruitfully cooperate with humans, and in designing  technical solutions to address and solve the application challenges.

In fact, we claim that a paradigm shift in the research on GRSs should be promoted, moving from the bare design of algorithms that aggregate group members' preferences and generate recommendation for selected items to a more holistic combination of functions for group decision making support. This support should unfold during the natural interaction of the group members, empowering the group members with better understanding of the decision making problem, helping them to better assess the other members' needs, and favoring respect and agreement. 

This research area must be approached in a multidisciplinary way. Moreover, we believe that it can flourish when the research community will be able to identify key application areas where group recommendations are required especially for decisions that are strategic and high risk. This will better motivate the usage of such systems. The emerging area of RSs for society development and sustainability~\citep{BorattoFL024} is proposed to be a valuable and challenging arena for the CAJO structure described here. The exploitation of the CAJO structure can enable a fairer and more productive collaboration between stakeholders.

Many open issues should be tackled to make progresses. The research community should identify knowledge sources and techniques that can enhance the full group dynamic support, from goal definition to final decision-making step. The process itself, i.e., how a group tackles a decision-making problem, has to be considered as important as the outcome. A successful process must balance the contributions of the group members and evolve in a rational way. During the process the GRS should play the role of a sensible group member that can detect the other members’: personalities, goals, satisfaction criteria, and knowledge both of the items and of the other group members beliefs. Focusing on the process also means to build support functionalities that dynamically update the representation of the group members, their preferences and goals, as they evolve while interacting with them. 

Moreover, group members have important relationships, e.g., family or work, that influence their actions and goals, which the GRS should recognize and leverage. We have stated in this position paper that the GRS should operate under diverse roles: arbiter of the process, oracle for information and explanation generation, judge making assessments and decision, and coach helping the user to improve the group performance, e.g., by giving appropriate feedback to group members and nudging them to adopt behavior that finally result in better decision outcomes.

Artificial intelligence and information retrieval technologies such as LLMs, video analysis and understanding, virtual reality presentation and simulation, are important means to further explore this research area. Finally, novel graphical user interfaces are required to organically aggregate the full range of functionalities of the system in collaboration with  the group.


\bibliography{sigirforum}
\end{document}